\documentclass[]{aastex631}
\usepackage{xcolor}

\begin{document}

\title{Halfway to the Peak: ice absorption bands at $z\approx0.5$ with JWST MIRI/MRS}

\author[0000-0002-1917-1200]{Anna Sajina}
\affil{Department of Physics and Astronomy, Tufts University, Medford, MA 02155, USA}

\author[0000-0001-8592-2706]{Alexandra Pope}
\affil{Department of Astronomy, UMass Amherst, Amherst, MA 01003, USA}

\author[0000-0002-8712-369X]{Henrik Spoon}
\affil{Cornell University, Ithaca, NY, USA}

\author[0000-0003-3498-2973]{Lee Armus}
\affil{California Institute of Technology, 1200 E. California Blvd., Pasadena, CA 91125, USA}

\author[0000-0002-4690-4502]{Miriam Eleazer}
\affil{Department of Astronomy, UMass Amherst, Amherst, MA 01003, USA}

\author[0000-0003-1748-2010]{Duncan Farrah}
\affil{UHawaii, Manoa, HI, USA}

\author[0000-0002-3032-1783]{Mark Lacy}
\affil{NRAO, Charlottesville, VA, USA}

\author[0000-0001-8490-6632]{Thomas Lai}
\affil{California Institute of Technology, 1200 E. California Blvd., Pasadena, CA 91125, USA}

\author[0000-0002-6149-8178]{Jed McKinney}
\affil{University of Texas at Austin, Austin, TX, USA}

\author[0000-0002-3158-6820]{Sylvain Veilleux}
\affil{Department of Astronomy and Joint Space-Science Institute, University of Maryland, College Park, MD 20742, USA}

\author[0000-0003-1710-9339]{Lin Yan}
\affil{California Institute of Technology, 1200 E. California Blvd., Pasadena, CA 91125, USA}

\author[0000-0002-5830-9233]{Jason Young}
\affil{Williams College \& SETI Institute, USA}

\begin{abstract}
This paper presents the first combined detections of CO$_2$, CO, XCN and water ices beyond the local Universe. We find gas-phase CO in addition to the solid phase CO. Our source, SSTXFLS J172458.3+591545, is a $z=0.494$ star-forming galaxy which also hosts a deeply obscured AGN. The profiles of its ice features are consistent with those of other Galactic and local galaxy sources and the implied ice mantle composition is similar to that of even more obscured sources. The ice features indicate the presence of a compact nucleus in our galaxy and allow us to place constraints on its density and temperature ($n>10^5$cm$^{-3}$ and $T=20-90K$). We infer the visual extinction towards this nucleus to be $A_V\approx6-7$. An observed plot of $\tau_{Si}$ vs. $\tau_{CO2}/\tau_{Si}$ can be viewed as a probe for both the total dustiness of a system as well as the clumpiness of the dust along the line of sight. This paper highlights the potential of using {\sl JWST} MIRI spectra to study the dust composition and geometric distribution of sources beyond the local Universe.  
\end{abstract}


\section{Introduction}

The mid-IR ($\sim$3-30\,$\mu$m) spectral regime is rich in ISM diagnostic features such as fine structure lines of atoms and ions as well as ro-vibrational bands of molecules in the gas or solid phase \citep[see][]{SpinoglioMalkan1992,Boogert2015,Sajina2022}. This makes it an excellent regime to explore the physical characteristics of the ISM. Ices including water, CO, CO$_2$ and others are expected to form on the surfaces of dust grains within cold, dense molecular environments including dense cores of molecular clouds, the dusty envelopes of YSOs, and protoplanetary disks \citep[see][for a review]{Boogert2015}.  Observations of ices through mid-IR absorption features has a long tradition in Galactic and Solar System studies \citep[e.g.][]{Oberg2008,Boogert2008,Oberg2011,Noble2013,Shirahata2013,Najita2018,Reach2023,McClure2023}. Laboratory studies and simulations have been critical in interpreting these data \citep[see e.g.][]{Rocha2021,Nietiadi2022,Dartois2022}. 

These observational, theoretical and laboratory studies suggest ice mantle formation through absorption of simple species from the gas-phase \citep[e.g.][]{vanScheltinga2022}; increasingly complex molecule formation on the surfaces of dust grains; and eventual desorption of said molecules back into the gas phase.  Ice mantle growth is likely to happen in different stages starting with the appearance of H$_2$O and CO$_2$ ices at low temperatures ($T<20K$) and moderately high densities ($n>10^4/cm^3$) and optical extinctions ($A_V>1.5$); followed by CO freeze-out at higher densities ($n>10^5/cm^3$) and extinctions ($A_V>3$), a phase often accompanied by ice species involving a C$\equiv$N bond; and finally even more complex chemistry including formation of methanol ice at $A_V>9$. Thus the precise ice composition provides information about the local conditions especially density. Similarly, some features, such as the 4.67\,$\mu$m CO ice band, are actually a complex of multiple species with varying sublimation temperatures. The relative strength of these sub-components is then a probe of the ice temperature \citep{Boogert2015}. More broadly, the presence of ice mantles is important for bulk dust properties such as albedo and size distribution. For example, icy mantles are believed to facilitate grain coagulation because when two ice-covered grains collide, there is partial melting in the interface zone which helps them stick together \citep{Nietiadi2022} leading to more efficient grain growth. The profiles of the ice features themselves have been suggested to be a sensitive probe of the grain sizes \citep{Dartois2022}. All of these imply the study of ice mantles on dust grains can provide a new probe of the chemistry and physical conditions especially in the dustiest environments where we have few other probes.  

The still relatively rare extragalactic observations of ice features are indeed limited to galaxies and AGN that have very dusty nuclei. For example, {\sl ISO} data showed various post CO freeze-out ice features in the dusty starburst and AGN NGC4945 \citep{Spoon2000} and the deeply obscured AGN, NGC4418 \citet{Spoon2001}. Ice features are absent in starburst and AGN that are not heavily dust obscured (e.g. Seyferts) \citep{Spoon2002},  but are common in local ULIRGs \citep{Imanishi2008}, especially for sources with buried AGN. To understand why, consider that most local ULIRGs, specifically those with cold {\sl IRAS} colors, show deeper silicate absorption features than typical starburst galaxies \citep[e.g.][]{Desai2007,Spoon2022}. The silicate absorption features are indicative of a large total column density of dust, but are not sensitive to the relative contribution of diffuse vs. dense cloud dust. There are some observations suggesting the dust dominating the silicate absorption to be on galaxy scales as opposed to concentrated in dusty nuclei \citep[e.g.][]{Lacy2007,Goulding2012}. However, build-up of ice mantles requires high physical density.  Therefore, unlike deep silicate absorption features alone, ice absorption features imply the presence of a physically compact (high density) nucleus, as discussed in \citet{Lai2020}. This helps break some of the uncertainties in where the dust is located in dust obscured systems \citep[e.g.][]{Marshall2018}. Ice absorption features can thus be complementary to ALMA observations of emission lines (e.g. from HCN) probing high density gas \citep[e.g.][]{Tokuda2014,vanderWiel2019,Murillo2022} including for galaxies \citep[see e.g.][and references therein]{French2023,Lin_2024}. However, these gas features are weak and it is hard to use them to probe the dense media of galaxies at higher redshifts. But even with {\sl Spitzer}, \citet{Sajina2009} were able to detect water ice absorption at $z\sim$2 and, with their limited sample, argued that there is little evidence for redshift evolution in the ice mantle properties of local ULIRGs and their cosmic noon analogues. Pre-{\sl JWST}, the most distant detection of the 4.27\,$\mu$m CO$_2$ and 4.67\,$\mu$m CO features (in the gas state) was for IRAS F00183-7111 at $z=0.327$, based on early {\sl Spitzer} IRS data \citep{Spoon2004}.
These rare studies, with typically poor spectral resolution, have not allowed us yet to realize the full potential of using ice features to gain further insights into the nature of heavily dust obscured galaxies.

With the advent of the {\sl JWST} Mid InfraRed Instrument \citep[MIRI;][]{Rieke2015}, we are poised for a dramatic increase in our ability to study the icy universe from Solar System bodies and Galactic objects out to distant dust obscured galaxies and AGN \citep[e.g.][]{McClure2023}. In particular, the {\sl JWST} ERS program 1328 (Co-PIs: L. Armus and A. Evans) has already increased the number and quality of ice feature detections among nearby galaxies \citep[e.g.][]{Lai2022,Rich2023}. The spectral resolution of the MRS is sufficient to not only detect the species in the solid phase, but also the gas phase \citep{Buiten2023} allowing gas excitation diagnostics similar to those using the mm-wave CO transitions detected by for example ALMA. As we increase the sample of objects with MIRI spectra we can look at population trends which can potentially reveal differences between heavily obscured AGN and starbursts. For example, we already have some evidence that dust grain sizes may be larger leading to greyer attenuation curves in heavily obscured galaxies and AGN \citep{LoFaro2017,Shao2017,Roebuck2019}. Grey attenuation curves out to the near-IR (implying micron-sized or larger grains) have significant implications including under-estimated galaxy stellar masses \citep{LoFaro2017}. However, existing studies focusing on the attenuation curves are degenerate in whether such grey attenuation is due to grain growth or the relative stars-dust geometry \citep[see][for a discussion]{Roebuck2019}. The profiles of mid-IR ice features can be invaluable in breaking this ambiguity as they are independently sensitive to grain growth \citep{Dartois2022} as well as being sensitive to physical density rather than only the dust column density.  Lastly, correctly accounting for ice absorption can be critical to the analysis and interpretation of the mid-IR continuum and diagnostic features such as PAHs \citep[see][]{Lai2020,Lyu2022,Sajina2022}. 

In this paper, we present the first detection of ice features due to CO$_2$, XCN, CO and water ice in a $z\approx0.5$ source. This redshift corresponds to roughly 5 billion years ago, i.e. this is the first detection of these ice features at an epoch sufficiently removed from the local Universe that we might expect evolutionary trends to be detectable. Incidentally, in studying the dust ice mantle properties at this redshift, we are also roughly probing the epoch of formation of our own Solar System. 
Our source has strong PAH features indicative of star-formation, but also has a deeply obscured AGN evidenced by the detection of the highly ionized neon features [NeV] and [NeVI] (see Young et al. 2024 in prep. and Pope et al. 2025 in prep.), as well as strong silicate absorption already seen in its {\sl Spitzer} IRS spectrum \citep{Lacy2007}. We present the {\sl JWST} MIRI/MRS spectrum analysis of this source and compare the ice features therein with observations of nearby galaxies including with recent {\sl JWST} data. This analysis helps us better understand the characteristics of the dust within our target and point to the potential of {\sl JWST} MIRI spectra in the study of deeply obscured sources beyond the local Universe.  

\section{Data}
\subsection{The source} \label{sec:source}
SSTXFLS J172458.3+591545 is a dusty starburst-AGN composite with an optical spectroscopic redshift of $z=0.494$ 
\citep{Lacy2007}. It was selected as an AGN on the basis of its IRAC color-color diagram and tentatively classified as a Type 2 AGN through optical spectroscopic diagnostics \citep{Lacy2007}. It was followed-up with both {\sl HST} imaging and {\sl Spitzer} IRS in \citet{lacy07_hst}. The latter revealed strong PAH emission, suggesting star-formation dominance. However, its relatively shallow 14/5.5\,$\mu$m slope ($f_{\nu,14}/f_{\nu,5.5}\approx2$) coupled with significant silicate absorption, $S_{sil}=ln(F_{obs,9.7}/F_{cont,9.7})=-1.5$ based on the {\sl Spitzer} IRS spectrum, are consistent with the presence of a buried AGN \citep[e.g.][]{Spoon2007,Marshall2018}. Using the spectral decomposition method of \citet{Kirkpatrick2015}, the AGN contributes 23\% of the mid-IR light in this source. 

Note that in the $S_{sil}$ definition there is no negative sign in the equation, hence the derived silicate absorption feature depth is negative. The reason for this convention is that the silicate feature can appear both in absorption and in emission. In later sections, where we compare this feature with the depth of other absorption features in the spectrum, we will switch to $\tau_{sil}$ which is related to $S_{sil}$ as $\tau_{sil}=-S_{sil}$ for consistency with the literature on ice features where the measured optical depths are always positive. 

This source is one of eight targetted by our {\sl JWST} GO1 program ``Halfway to the Peak: A Bridge Program To Map Coeval Star Formation and Supermassive Black Hole Growth" (PID1762; PIs Pope, Sajina, Yan). The targets focus on the $z\sim0.5-0.6$ epoch and span a range in the relative fraction of their mid-IR luminosity that is due to star-formation or an AGN. SSTXFLS J172458.3+591545 (FLS2 within our program) is the source with the second lowest AGN fraction, see above. However, it has the deepest silicate absorption within our sample. 

\subsection{JWST MIRI/MRS data}

Observations were carried out in July 2022 with the {\sl JWST} MIRI's Medium Resolution Spectrometer \citep[MRS; ][]{Wells2015,Wright2023} as part of the GO1 program \#1762 (see above).  For full details of the observations and data reduction see Young et al. (in prep). Here we give only a brief summary. The source was observed for 2220s in each of the three MRS grating settings (short, medium and long). Each grating setting observes simultaneously in the four MRS channels such that the combined data consists of 12 sub-bands that span 4.9-27.9\,$\mu$m. The data were reduced and assembled into spectral cubes using a customized notebook. Our notebook calls the standard {\sl JWST} Science Calibration Pipeline 1.9.6 \citep{Bushouse2023}, but includes custom steps focusing on background subtraction. The key is that after source masking, the stage 1 rate files are median combined on a per detector/grating setting basis to generate a master background. Due to time variable backgrounds, only sources observed within a month of each other included in a particular master background. As demonstrated in Young et al. (in prep), our customized pipeline achieves consistently better results than the standard pipeline alone with the achieved 1\,$\sigma$ RMS noise for each of the 12 MRS sub-bands essentially consistent with pre-flight expectations \citep{Glasse2015}. As a guideline the 1\,$\sigma$ RMS noise is $\approx1-2\times10^{-21}$W/m$^2$ for the shorter wavelength sub-bands (corresponding to MRS channels 1 and 2), but is more than an order of magnitude worse in the reddest sub-bands (corresponding to MRS channel 4). 

\begin{figure*}[h]
\centering
\includegraphics[scale=0.65,clip=true]{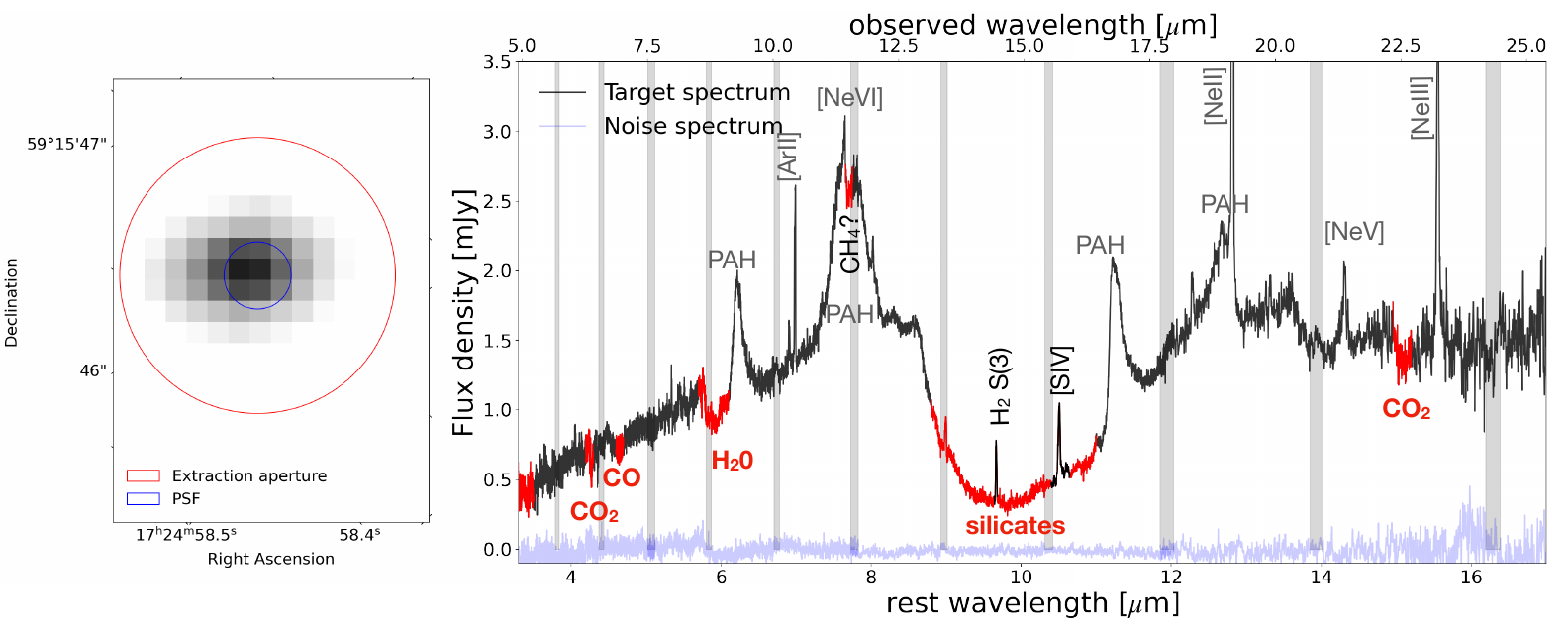}
\caption{{\it Left:} The collapsed channel 1 MRS cube of FLS2. We overlay the JWST MIRI PSF at 5.6\,$\mu$m as well as the aperture with a radius of 0.85$\arcsec$ (based on the PSF in channel 4) that we use for our total spectrum extraction. {\it Right:} The JWST MIRI/MRS spectrum of SSTXFLS J172458.3+591545 from the $r$=0.85$\arcsec$ aperture extraction. The blue/purple spectrum is the noise spectrum extracted from an equal sized aperture offset from the source. The grey vertical stripes mark the overlap regions between the 12 MRS sub-bands. We label the more prominent emission features including [NeVI] and [NeV] which show the presence of an AGN. The regions colored in red mark molecular absorption bands as labeled. 
\label{fig:full_spectrum}}  
\end{figure*}

\subsection{Spectral extraction \label{sec:spectra}}
Figure\,\ref{fig:full_spectrum}{\it left} shows the channel1 collapsed image of our source, FLS2. Note that the spatially-resolved features are discussed further in Young et al. (in prep.). We extract the 1D spectrum in two apertures. One is our ``total" aperture which has a radius of 0.85$\arcsec$ and is designed to encompass the full light of the source even at the longest wavelengths since this radius is approximately the FWHM of the PSF of the longest wavelength channel (channel4). The other extraction uses a PSF aperture where for each channel the extraction aperture corresponds to the maximum PSF within that channel. Note that the PSF aperture in channel 1 (radius of 0.25$\arcsec$) corresponds to a physical radius of 1.5kpc at the redshift of our source. As shown in Figure\,\ref{fig:full_spectrum} the source extends beyond this PSF aperture.

Figure\,\ref{fig:full_spectrum}{\it right} shows the MIRI/MRS 1d spectrum of SSTXFLSJ172458.3+591545 from the 0.85\,$\arcsec$ aperture.  Here we see no offsets between the channels since the same large aperture is used throughout the spectrum. The same figure for the PSF extraction shows steps between the channels. Figure\,\ref{fig:full_spectrum} also shows the noise spectrum extracted from the same size aperture, but spatially offset relative to the source. The mean of the noise spectrum is -0.02\,mJy with a standard deviation of 0.06\,mJy.  The continuum is detected throughout the spectrum with a minimum SNR of $\approx$10. 

We mark in Figure\,\ref{fig:full_spectrum} the key emission and absorption features. The methane ice feature at 8\,$\mu$m is with a question mark due to the difficulty of distinguishing from the detailed structure of the 7.7\,$\mu$m PAH complex.  We also note the potential 15\,$\mu$m CO$_2$ absorption feature. Since this part of the spectrum is quite noisy, this feature however is not significantly detected. Therefore we only focus on the 4.27\,$\mu$m $^{12}$CO$_2$ feature in this paper.  

\begin{figure*}[h]
\centering
\includegraphics[scale=0.35,clip=true]{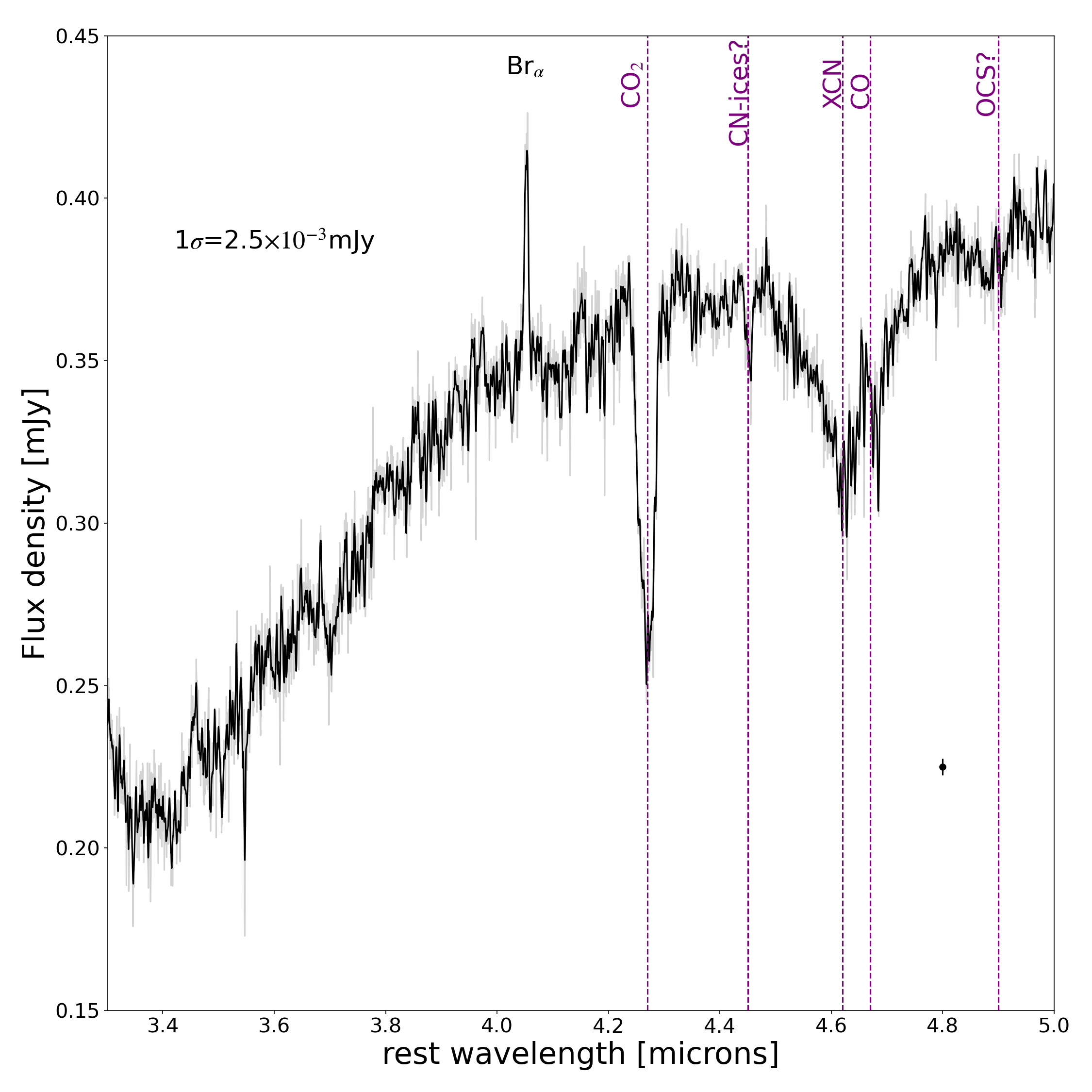}
\caption{The MIRI/MRS channel 1 spectrum of our source using a PSF-matched extraction aperture. The greyscale shows the raw spectrum, while in black we show the spectrum after applying a Savitzky-Golay smoothing filter with a window size of 10. The adopted optical spectroscopic redshift ($z=0.494$) \citet{Lacy2007} fully matches the narrow Br$\alpha$ line as shown. The rest-frame wavelengths of the marked ice absorption features are based on \citet{Boogert2015} and \citet{Nazari2024} for the potential CN-ices feature. Finally, a representative 1\,$\sigma$ uncertainty for the smoothed spectrum is listed. 
\label{fig:ch1_spectrum}}  
\end{figure*}

Figure\,\ref{fig:ch1_spectrum} shows the PSF extracted channel 1 spectrum where we label the detected emission and absorption features. We note that the same ice features (with roughly the same strength) are present in both the total ($r$=0.85$\arcsec$) and the PSF extracted spectra, but are much less noisy in the later. To further reduce the noise we applied a Savitzky-Golay smoothing filter with window size 10\footnote{Unlike a simpler moving average smoothing, this filter computes a polynomial within the given window and uses that to compute the appropriate value at the window mid-point}. Figure\,\ref{fig:ch1_spectrum} shows both the raw and the smoothed spectra showing that the shape and strength of the absorption features are unaffected by the smoothing process. The errorbar on the side is representative of the noise within this binned window\footnote{ This is the standard deviation of the noise spectrum in channel 1 divided by the square root of the smoothing window size.}. 

The most prominent ice absorption features in Figure\,\ref{fig:ch1_spectrum} are due to $^{12}$CO$_2$, XCN and CO ice. We also detect a weaker absorption feature at 4.45\,$\mu$m. Laboratory studies show absorption in this regime is associated with a multitude of nitrogen-bearing ices \citep[e.g.][]{Moore_2010,Rachid2022} especially due to CN-stretching. Composite absorption features at $\approx$4.45\,$\mu$m are also seen in the {\sl JWST} spectra of several YSOs \citep{Nazari2024} and are attributed to the nitrogen-bearing complex organic molecules methyl cyanide CH$_3$CN and ethyl cyanide C$_2$H$_5$CN. Indeed, \citet{Nazari2024} is the first detection of these molecules in the solid state in astronomy. Our feature is narrower (spanning 4.44-4.46\,$\mu$m) and is centered on 4.455\,$\mu$m which is closest to the C$_2$H$_5$CN peak at 4.452\,$\mu$m. We note that the 4.45\,$\mu$m to 4.62\,$\mu$m (XCN) peak optical depth ratio in our galaxy is about $\sim$20\,$\times$ larger than the YSOs studied in \citet{Nazari2024}. In addition, there is broad absorption at $\sim$4.9\,$\mu$m which could be due to the 4.90\,$\mu$m C-O stretch of OCS ice \citep{Boogert2015}. 
The $^{13}$CO$_2$ feature at 4.39\,$\mu$m is not detected. We also looked for the dense medium tracer CH$_3$OH at 3.53\,$\mu$m, but it is not detected. There is evidence for the broad hydrocarbon feature at 3.4\,$\mu$m, but we do not further pursue it due to the lack of spectral coverage below 3.3\,$\mu$m. Lastly, we could not identify the weak absorption at 3.7$\mu$m. For the ice feature analysis in Section\,\ref{sec:analysis}, we use this PSF-extracted and smoothed spectrum.  

\begin{figure*}[h!]
\centering
\includegraphics[scale=0.4,clip=true]{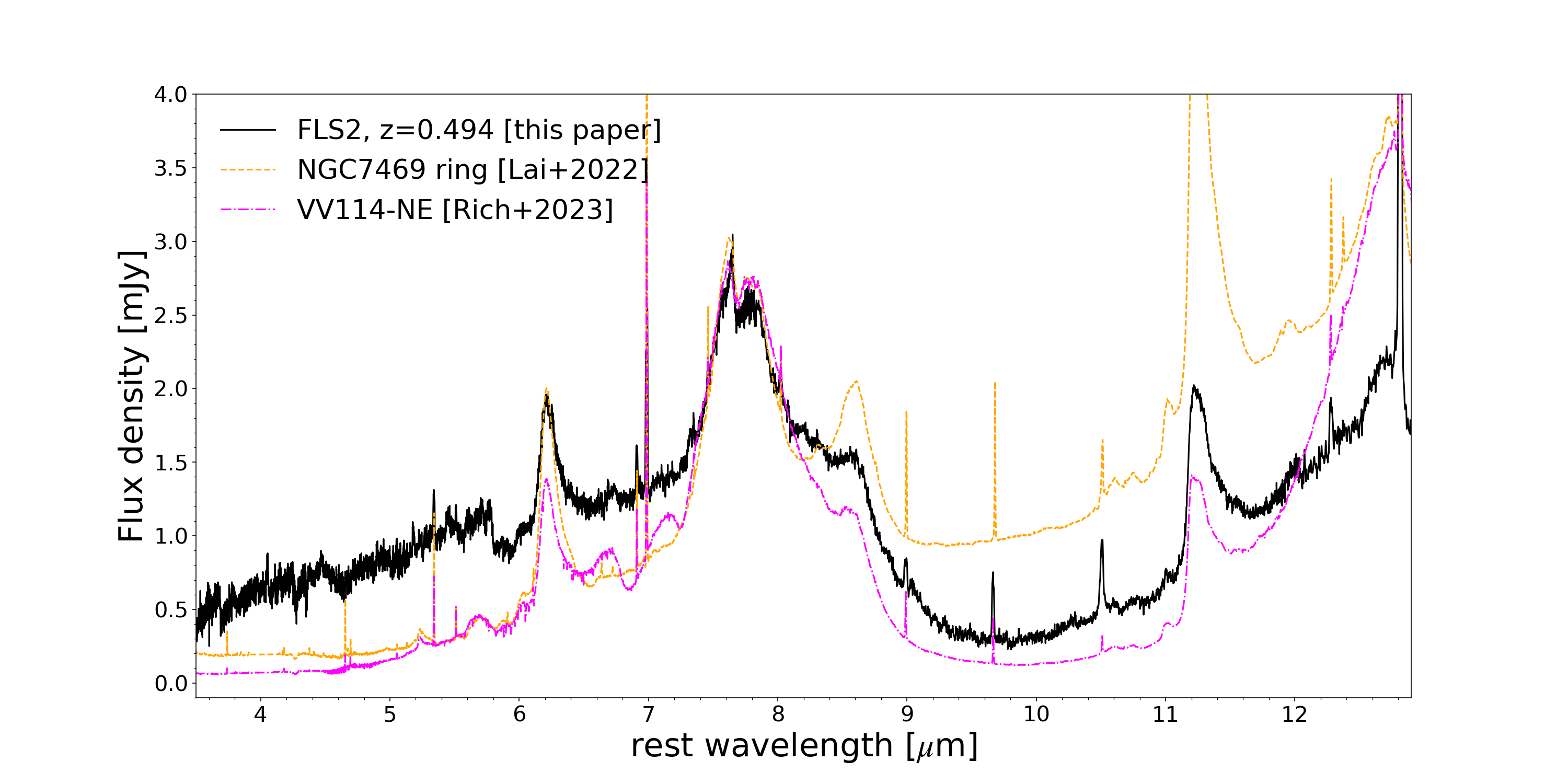}
\caption{Our MIRI/MRS spectrum compared with two JWST NIRSpec+MIRI spectra from nearby galaxies (scaled to match our spectrum at $\approx$7.7\,$\mu$m). The two local galaxy comparison spectra contain ice absorption features and show 9.7\,$\mu$m silicate feature depths that are both shallower and deeper than our source based on their silicate strengths (see text for details). Note that our source has the strongest ``hot dust" continuum consistent with the presence of a buried AGN. 
\label{fig:compare_goals}}  
\end{figure*}

\section{Analysis}
\label{sec:analysis}

\subsection{Comparison with the JWST mid-IR spectra of nearby galaxies}
\label{sec:comparison}

Our first step is to compare the full mid-IR spectrum of our $z\sim0.5$ galaxy with two recently published JWST spectra of local sources that also show ice absorption features. Figure\,\ref{fig:compare_goals} shows our JWST spectrum overlaid with those of the ring of NGC7469 \citep{Lai2022} and the North East (NE) region in VV114 \citep{Rich2023}.  The overall mid-IR spectrum of our source shows a stronger hot dust continuum component around 5\,$\mu$m consistent with a buried AGN. The AGN is confirmed by the presence of [NeV] and [NeVI] seen in Figure\,\ref{fig:full_spectrum}{\it right}. 

Our source has a silicate absorption feature that is deeper than the NGC7469 ring spectrum, but shallower than VV114-NE. Using the common definition\footnote{This definition constructs a linear continuum between 5.5 and 12\,$\mu$m and estimates the silicate feature depth at 9.7\,$\mu$m as $S_{sil}=ln(F_{obs}/F_{cont})$.} of the observed silicate feature depth \citep[e.g.][]{Marshall2018} we find $S_{sil}=-1.3$ for our source\footnote{With a slightly different continuum treatment, the IDEOS database finds $S_{sil}\approx-1$ \url{http://ideos.astro.cornell.edu/cgi/single.py?ideosid=14017024_0}.}.  The same procedure gives $S_{sil}=-0.5$ for NGC7469-ring and $S_{sil}=-2.75$ for VV114-NE\footnote{\citet{Rich2023} calculate $S_{sil}$=-2.45 for VV114-NE which is reasonably consistent with our value given the dependence on the details of how the continuum is defined.}.

Figure\,\ref{fig:compare_goals} shows that, despite its intermediate silicate absorption strength, our source has the strongest water ice absorption at 6\,$\mu$m along with much more prominent CO+XCN absorption complex at 4.67\,$\mu$m. The CO$_2$ feature is present in all three spectra, as we will show more clearly in Section\,\ref{sec:opt_depth}. Unlike the deepest $S_{sil}$ spectrum, VV114-NE, our source lacks the aliphatic C-H absorption features at 6.85\,$\mu$m and 7.25\,$\mu$m. These features are also seen in deeply obscured spectra towards the Milky Way Galactic Center \citep{Chiar2000,Chiar2002}, and should not be confused with the 6.85\,$\mu$m feature seen in e.g. W33A which is attributed to CH$_3$OH and NH$_4+$ ices \citep{Boogert2015}. Conversely, NGC1377 is an example of a deeply obscured nucleus without either water ice or C-H absorption features \citep{Roussel2006}. The extensive spectroscopic database of local {\sl Spitzer} spectra, IDEOS, includes both silicate and C-H optical depth measurements but finds no correlation between the two \citep{Spoon2022} except that C-H features do not appear in sources without silicate absorption. 

\begin{figure*}[h!]
\centering
\includegraphics[scale=0.22,clip=true]{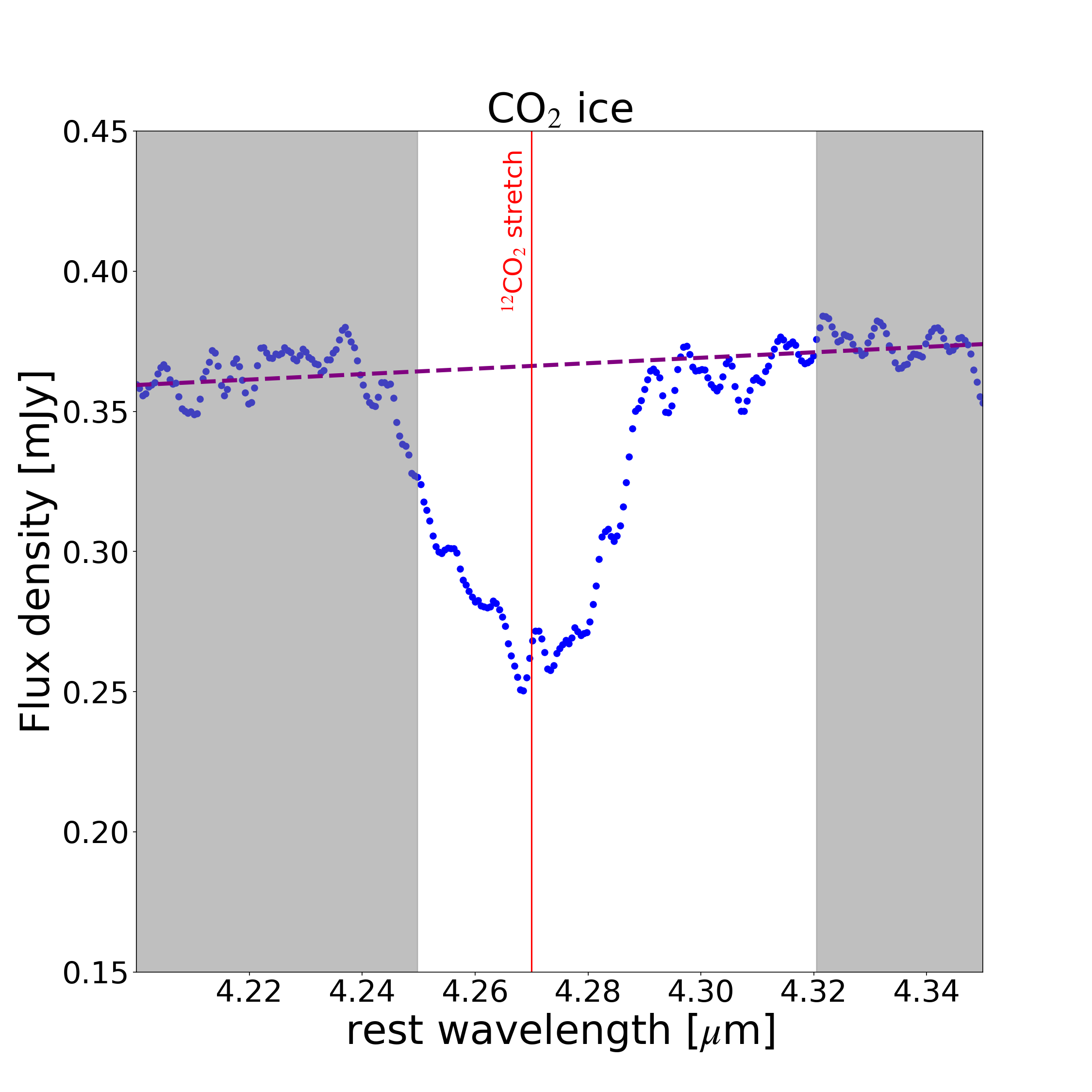}
\includegraphics[scale=0.22,clip=true]{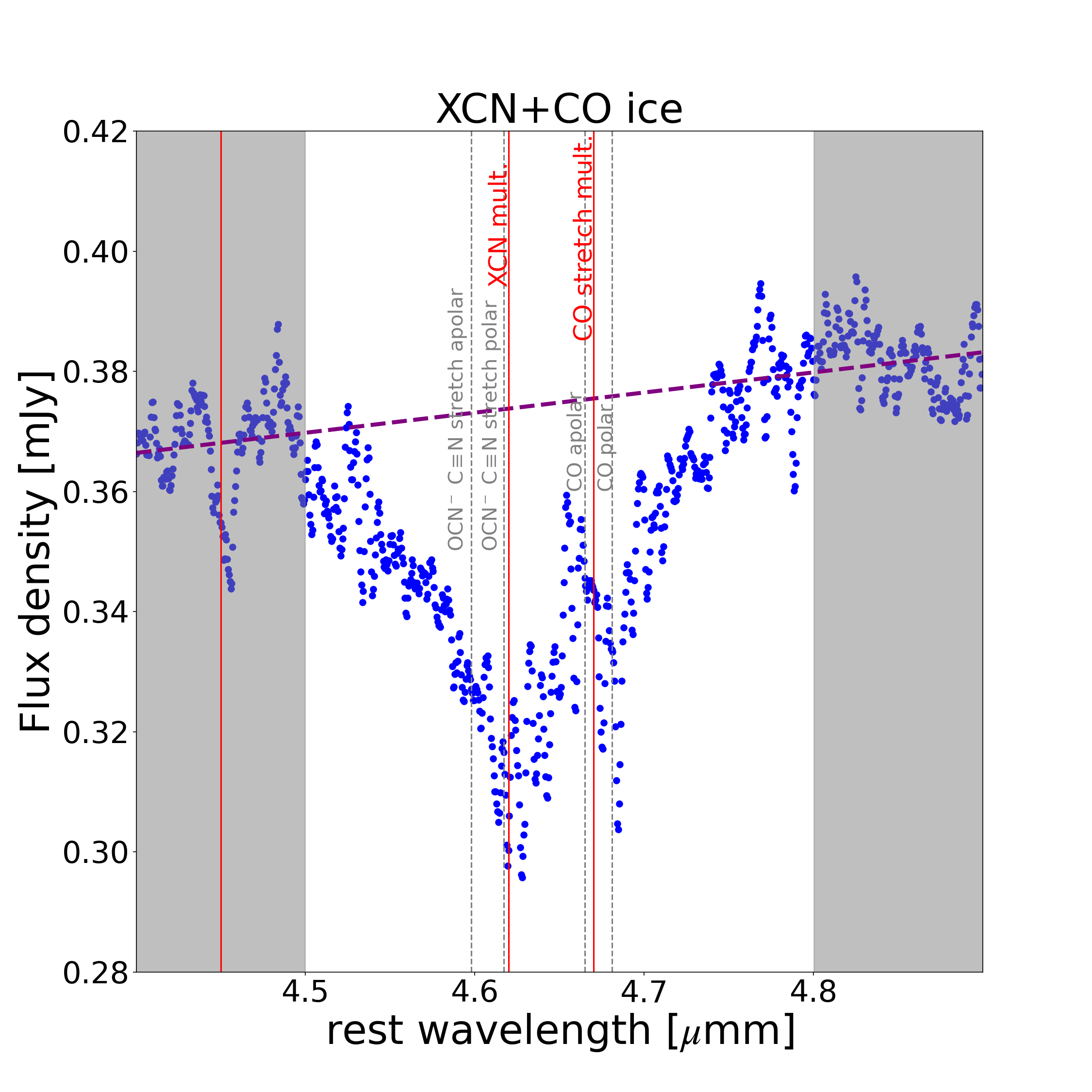}
\caption{Zoom-in spectra around the CO$_2$ ({\it left}) and XCN+CO ({\it right}) features. The MRS spectra, smoothed as discussed in Figure\,\ref{fig:ch1_spectrum} are in blue. The grey bands show the areas used in the linear continuum fit. The thick purple dashed lines are the resulting continua. The vertical lines denote the positions of known absorption features, as labeled, based on Table\,1 in \citet{Boogert2015}. The solid red lines are the features used in the optical depth determinations. The "XCN" and "CO" features are both multi-component features. Their main (polar/apolar) contributors are indicated by the vertical grey dashed lines. In both cases, the polar component dominates. 
\label{fig:zoom_spectra}}  
\end{figure*}

\begin{figure*}[h!]
\centering
\includegraphics[scale=0.4,clip=true]{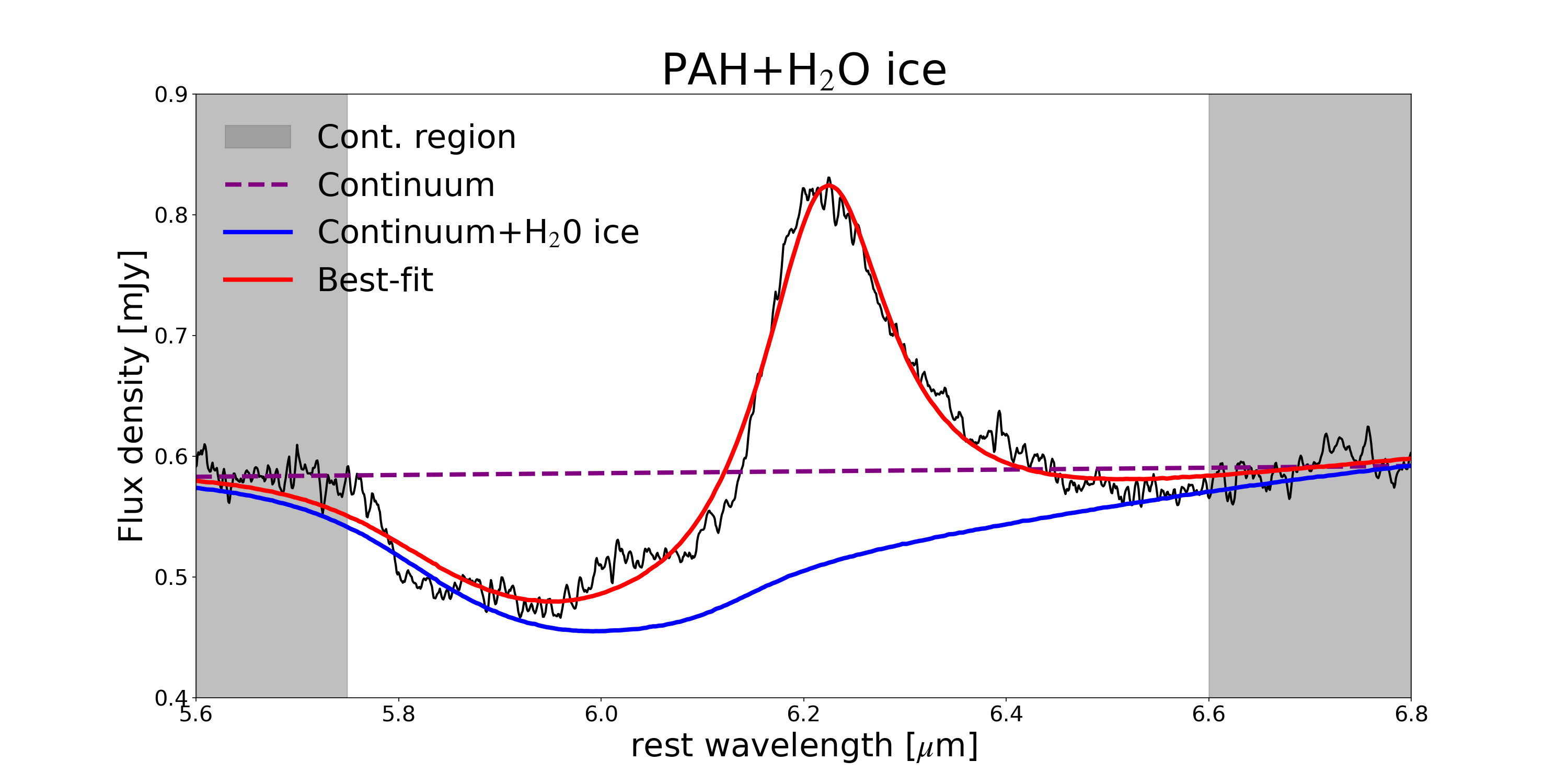}
\caption{A zoom-in around the rest-frame 6.0\,$\mu$m water ice and 6.2\,$\mu$m PAH features. 
As in Figure\,\ref{fig:zoom_spectra}, the continuum model is the dashed purple line which is fit to the grey shaded regions and is used for the optical depth determination. The combined continuum+ice+PAH (Equation\,\ref{eq:icepahmodel}) best-fit model is shown in red.
The best-fit continuum with water ice absorption only is shown in blue. The water ice absorption profile is based on the 10\,K laboratory spectrum of \citet{Gerakines1995}. 
\label{fig:zoom_water}}  
\end{figure*}

\subsection{Continuum determination around the ice absorption features}
\label{sec:cont_determ}

Figure\,\ref{fig:zoom_spectra} shows a zoom-in of the MRS spectrum around the CO$_2$ and CO absorption features where simple linear continua are fit within spectral windows free from known emission or absorption features. Figure\,\ref{fig:zoom_spectra} also marks the locations of the known ice bands based on \citet{Boogert2015}. Note that the redshift here is adopted from the narrow emission lines (e.g. Br$\alpha$ seen in Figure\,\ref{fig:ch1_spectrum}) with no attempt to center the broad absorption features onto the nominal ice feature wavelength. Ice features profiles are known to vary from source to source depending on factors such as the precise composition of the ice mantles as well as the physical properties of the dust grains \citep{Gerakines1995,Boogert2015,Dartois2022}.    

Figure\,\ref{fig:zoom_water} shows a zoom-in of the MRS spectrum around the rest-frame 6.0\,$\mu$m which includes the water ice feature as well as the 6.2\,$\mu$m PAH feature. The absorption feature here is in fact a complex of multiple potential contributors including the C-O stretch in H$_2$CO, the NH$_3$ bend, the C-C stretch in PAH molecules (which can be seen in absorption as well as emission), and finally the H$_2$O bending mode \citep{Boogert2015}. This complex nature likely accounts for the somewhat uneven profile of this feature with noticeable bumps and troughs even outside the PAH feature. While avoiding much of this complexity, we can capture the main aspects of the spectrum using a simple empirical model given by:

\begin{equation}
    F_{\lambda,model}=F_{\lambda,cont}*exp(-\tau_{H_2O}*A_{\lambda,water})+F_{\lambda,PAH}
    \label{eq:icepahmodel}
\end{equation}
where $F_{\lambda,cont}$ is a linear continuum, $A_{\lambda,water}$ is the optical depth profile of the 6.0\,$\mu$m water feature based on laboratory measurements at 10K \citep{Gerakines1995}, and $F_{\lambda,PAH}$ is a single Lorentzian profile adopted for the PAH. The free parameters include two parameters for the linear continuum, the depth of the water ice feature, $\tau_{H_2O}$, and the amplitude, width and center of the PAH feature modeled by a Lorentzian profile. The observed PAH feature is slightly asymmetric which would suggest it too is subject to some extinction, if the intrinsic 6.2\,$\mu$m PAH feature were symmetric. However, this is likely not the case \citep[e.g.][]{Peeters2004,Draine2021,Spoon2022}, with the GOALS {\sl JWST} spectra for example finding that this feature is best described as a complex of three distinct features whose relative strength can give rise to a variety of profile shapes \citep[e.g.][]{Evans2022,Armus2023}.  
Figure\,\ref{fig:zoom_water} shows the best fit model. As discussed above the 6\,$\mu$m absorption feature is clearly more complex than the simple smooth pure water ice profile seen in the laboratory spectrum. There is noticeable excess absorption at $\approx$5.8\,$\mu$m and excess emission at $\approx$\,6.0\,$\mu$m relative to this simple best-fit model. This is reminiscent of the YSO spectra of this feature which often show multiple distinct absorption troughs \citep[see][]{Boogert2015}. Because of this complexity, when we use the depth of this feature to derive a water ice column density (see Section\,\ref{sec:results}), it is in fact an upper limit thereof.  

Lastly, we also explored a model where the PAH and continuum are both subject to the same water ice attenuation. However, the fit was better with the model form as given in Eq-n\,\ref{eq:icepahmodel}. It is conceivable that an independent attenuation term is needed for the PAH component, but as discussed above, this is difficult to disentangle from the uncertainties in the intrinsic PAH profile. We therefore preferred to adopt the simple model given in Eq-n\,\ref{eq:icepahmodel}.

\subsection{Optical depth profiles and column densities of the ice absorption features \label{sec:opt_depth}}

For the CO$_2$ and XCN+CO spectral windows, we determine the optical depth profiles using $\tau_{\nu}=ln(F_{\nu}/F_{\nu,cont.})$ where $F_{\nu}$ is the data in mJy and $F_{\nu,cont}$ are the best-fit linear continuum fits shown in Figure\,\ref{fig:zoom_spectra}. For the 6.0\,$\mu$m water ice feature, while our best-fit $\tau_{H_20}$ is reasonable, the full (PAH-subtracted) optical depth profile is too uncertain due to the strong overlapping PAH. It is therefore omitted in the discussion of the optical depth profiles in Section\,\ref{sec:results}, although we use the overall depth of the feature to place an upper limit on the water ice column density (see below).  

In order to obtain the peak and integrated optical depths of each feature, we fit single or double Gaussian profiles to these optical depth profiles. The results of these fits are given in Table\,\ref{table:derived_quantities}. The values we call $\tau_X$ where $X$ is CO$_2$, XCN, or CO always refer to the peak optical depth from these Gaussian fits. The value for $\tau_{H_2O}$ is based on the model fit using Equation\,\ref{eq:icepahmodel}. These ``peak" optical depth values are fairly analogous to the $S_{sil}$ values discussed in Section\,\ref{sec:comparison} in that they represent essentially the observed depth of these features relative to the linear continuum. From now on, the $S_{sil}$ values are referred to as $\tau_{sil}$ for consistency with this nomenclature. We also determine the integrated optical depths based on the standard formulation $\tau_{int,X}=\int\tau_{\nu,X}d\nu$ where $\nu$ is a wavenumber in units of [1/cm] \citep{Gerakines1995}. These integrated optical depths are used to derive column densities $N_{X}=\int\tau_{\nu,X}d\nu/A$ where $A$ is the molecular band strength. We adopt the band strengths of $1.1\times10^{-16}$,  $1.1\times 10^{-17}$, and $1.3\times10^{-16}$cm/molecule for $^{12}$CO$_2$, $^{12}$CO and water ice respectively, all based on \citet{McClure2023}.  We assume the bulk of the XCN feature can be attributed to the OCN$^-$ molecule and therefore adopt the band strength thereof ($1.2\times 10^{-17}$cm/molecule) from \citet{vanBroekhuizen2004}. The later assumption of course carries some uncertainty since this feature can include other ice species as well \citep{Boogert2015}. Table\,\ref{table:derived_quantities} shows all the profile shape fits, optical depths and column density determinations for the detected ice features in our source. 

\section{Results \label{sec:results}}
\subsection{The 4.27\,$\mu$m CO$_2$ feature}

\begin{figure*}[h!]
\centering
\includegraphics[scale=0.35,clip=true]{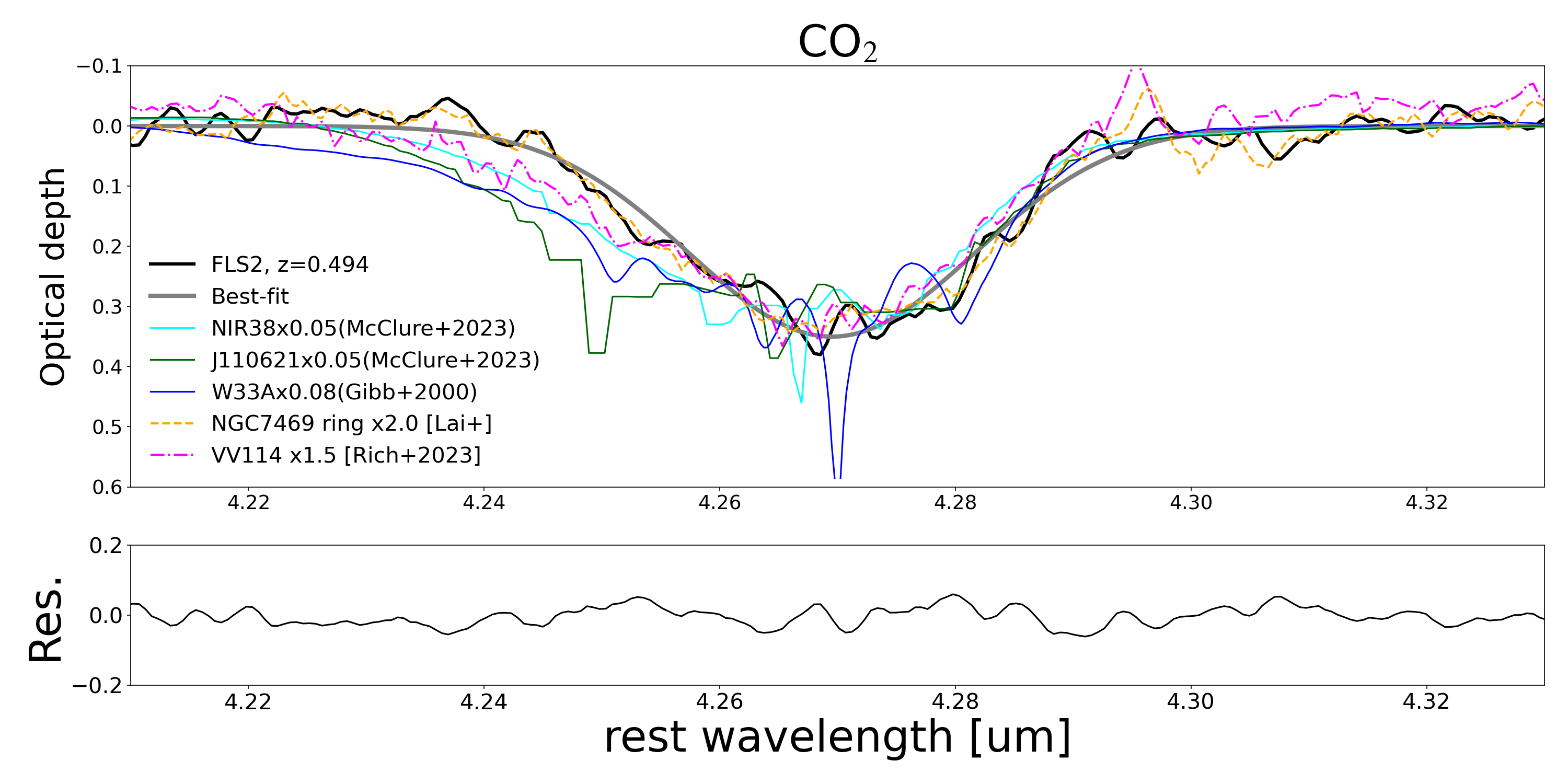}
\caption{The thick black curve is the optical depth profile of the 4.27\,$\mu$m CO$_2$ feature for FLS2. The best-fit Gaussian is the thick grey curve. The residual from subtracting this best-fit is shown at the bottom. Note that the standard deviation of this residual is comparable to the typical error in our spectra (after the Savitzky-Golay data filtering) suggesting there are no additional significant components. We also overlay the CO$_2$ optical depth profiles from several comparison spectra scaled as noted in the legend. The comparison spectra include NIR38, J11062 and W33 which are all extremely obscured Galactic targets with $A_V>50$, which accounts for their significantly deeper $CO_2$ features. {\bf Of the extragalactic sources,} our spectrum is particularly similar to that of the NGC7469 ring, though our CO$_2$ feature is about 2\,$\times$ deeper.  
\label{fig:opt_depth_co2}}  
\end{figure*}

Figure\,\ref{fig:opt_depth_co2} shows the 4.27\,$\mu$m $^{12}$CO$_2$ absorption feature seen in our source.  The weaker feature at 4.38\,$\mu$m due to $^{13}$CO$_2$ is not detected. We also overlay 
multiple Galactic and extra-galactic comparison spectra (scaled as noted) for comparison. The comparison spectra include NIR38, J11062 and W33 which are all extremely obscured Galactic sources with $A_V>50$, which accounts for their significantly deeper $CO_2$ features. It should be noted that the CO$_2$ features from these three heavily obscured sources are all saturated \citep[see][]{McClure2023} which means the exact depth and profiles are less certain. The CO$_2$ absorption profile for VV114 is very similar to all these heavily obscured sightlines but is not saturated. Our spectrum is particularly similar to that of the NGC7469 ring, though our CO$_2$ feature is about 2\,$\times$ deeper. Both optical depth profiles show an excess, i.e. less absorption, in their blue wing relative to the other comparison spectra, although this should be viewed with caution due to the saturated nature of many of these comparison spectra. 

Examining the CO$_2$ optical depth profiles for 14 Galactic sources, \citet{Gerakines1999} also note profile variation which they attribute to the relative strength of the dominant polar vs. apolar components.  The apolar component contributes additional absorption on the blue wing so our ``blue-excess" sources might be interpreted as having less of an apolar component. Similar ``blue-excess" in the $^{12}$CO$_{2}$ ice profile is observed in the AKARI spectra in multiple Galactic YSOs \citep{Noble2013}. Using radiative transfer models, \citet{Dartois2022} suggest that such asymmetric profiles can also arise as a result of grain growth, specifically when $a_{max}>$1\,$\mu$m. Further exploration of this effect is beyond the scope of this paper and requires careful radiative transfer modeling of our spectrum. However, we note that there are independent indications of grain growth in the vicinity of the AGN based on the optical to 9.7\,$\mu$m extinction ratio as well as the shift longward of the silicate emission feature in Type 1 AGN \citep{Shao2017}. 

\begin{figure*}[h!]
\centering
\includegraphics[scale=0.3,clip=true]{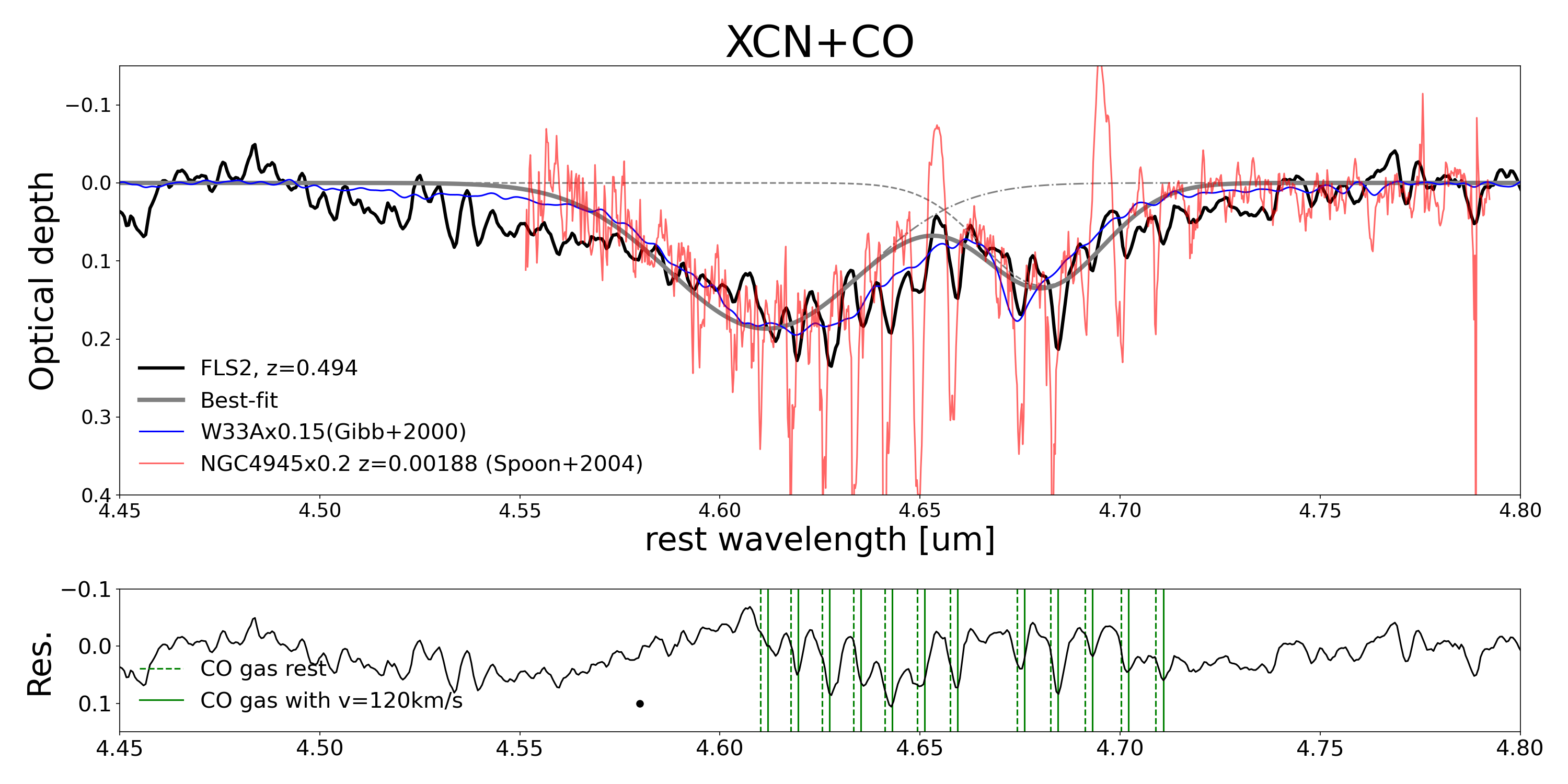}
\caption{The thick black curve is the optical depth profile of the 4.6\,$\mu$m XCN+CO absorption complex for FLS2. The grey curves are the single or double Gaussian best-fit models. Note that the NGC4945 spectrum has a slightly smaller spectral coverage. It also includes a couple of emission feature which we have not removed. The residual in the bottom shows hints of warm CO gas potentially shifted by 120\,km/s. 
\label{fig:opt_depth_co}}  
\end{figure*}

\subsection{The 4.6\,$\mu$m XCN+CO feature}
Figure\,\ref{fig:opt_depth_co} shows the optical depth plot of the 4.6\,$\mu$m XCN+CO complex. As discussed in Section\,\ref{sec:opt_depth}, we model this broad absorption by a double Gaussian. The narrower one at 4.67\,$\mu$m is due to the CO stretch mode while the broader one at 4.62\,$\mu$m is most commonly attributed to OCN$^-$, but is often referred to as  "XCN" due to its complex nature, dominated by the stretching of the C$\equiv$N bond\citep{Boogert2015}. Our spectrum looks remarkably similar to both the canonical obscured W33A spectrum as well as the NGC4945 spectrum from \citet{Spoon2004} -- the first known extragalactic observation of this feature. The NGC4945 spectrum is taken from the nucleus around the buried AGN in this system \citep{Spoon2004}. 

Note that the best-fit Gaussian for the CO absorption peaks at 4.681\,$\mu$m which is the exact wavelength for the polar component of the CO absorption feature \citep{Boogert2015}. As shown in the right-hand panel of Figure\,\ref{fig:zoom_spectra}, this feature has two components: an apolar one peaking at 4.673\,$\mu$m expected from pure CO ice, and a polar component at 4.681\,$\mu$m expected when the CO is part of $H_20$+CO ice mixture for example. 
The ratio of the polar to apolar component has been suggested as a temperature gauge since the apolar component sublimates at 20\,K whereas the polar component sublimates at 90\,K \citep{Boogert2015}. Our spectrum is consistent with temperatures of 20-90\,K.  

The NGC4945 spectrum includes both the broad ice features as well as superimposed sharp absorption features corresponding to the P+Q ro-vibrational spectrum of CO in the gas-phase \citep{Spoon2004}. The W33A spectrum shown does not indicate gas-phase CO due to its low spectral resolution \citep[$R\sim750$;][]{Gibb2000}. At higher resolving power ($R\sim37,000$) gas-phase CO lines are present \citep{Mitchell1990}. 

Recent {\sl JWST} of nearby galaxies show pure gas-phase CO instead \citep{Buiten2023,Gonzalez-Alfonso2024}, including the NGC7469 ring. Our spectrum shows both broad ice features and gas-phase CO. A recent spectrum of the local obscured AGN NGC3256 also shows a combination of ice and gas-phase absorption in this feature allowing for detailed studies of the physical conditions of the gas as well as a constraints on the $\alpha_{CO}$, the CO-to-H$_2$ conversion factor \citep{Pereira-Santaella2024}. 

Such a combination of ice absorption with superimposed CO gas absorption may in principle indicate the process of desorption of the solid-phase CO from the dust grain ice mantles back into the gas-phase. But considering that our spectrum corresponds to a large portion of the entire galaxy (given its redshift), it is also possible that the ice and gas components are not physically associated. In the residual spectrum (where the double Gaussian model is subtracted from the data), we overlay the rest wavelengths of the different modes of the ro-vibrational CO gas spectrum. We find reasonably good correspondence with the ``dips" in our residual indicative of the presence of gas-phase CO in addition to the CO ice. The positions of our dips however show a slight systematic redshift of $\approx$120\,km/s. Note that the MIRI/MRS velocity resolution in channel 1 is $\approx$80\,km/s.  If we assume this redshift is real, is it physically possible? It is unlikely that desorbing gas will have such a shift, hence this may argue in favor of an independent gas component. If true, then redshifted absorption would be indicative of an inflow. Outflows are seen as blueshifted absorption or P-Cygni profiles \citep[e.g.][]{Veilleux2013,Spoon2013}. A molecular gas inflow here would be similar to what we have in the center of the Milky Way for example \citep{Sormani2019}. Our galaxy also has the millimeter CO(2-1) line detected with IRAM which is discussed in the context of its molecular gas properties inferred from the mid-IR $H_2$ which indicate the presence of two temperature components (at $\approx$300K and $\approx$1000K) (Yan et al. in prep). Further discussion of how our tentative mid-IR CO gas detection relates to the mm-wave molecular gas constraints is left to Yan et al. (in prep.).

\subsection{Ice mantle composition analysis}
\begin{figure*}[h!]
\centering
\includegraphics[scale=0.65,clip=true]{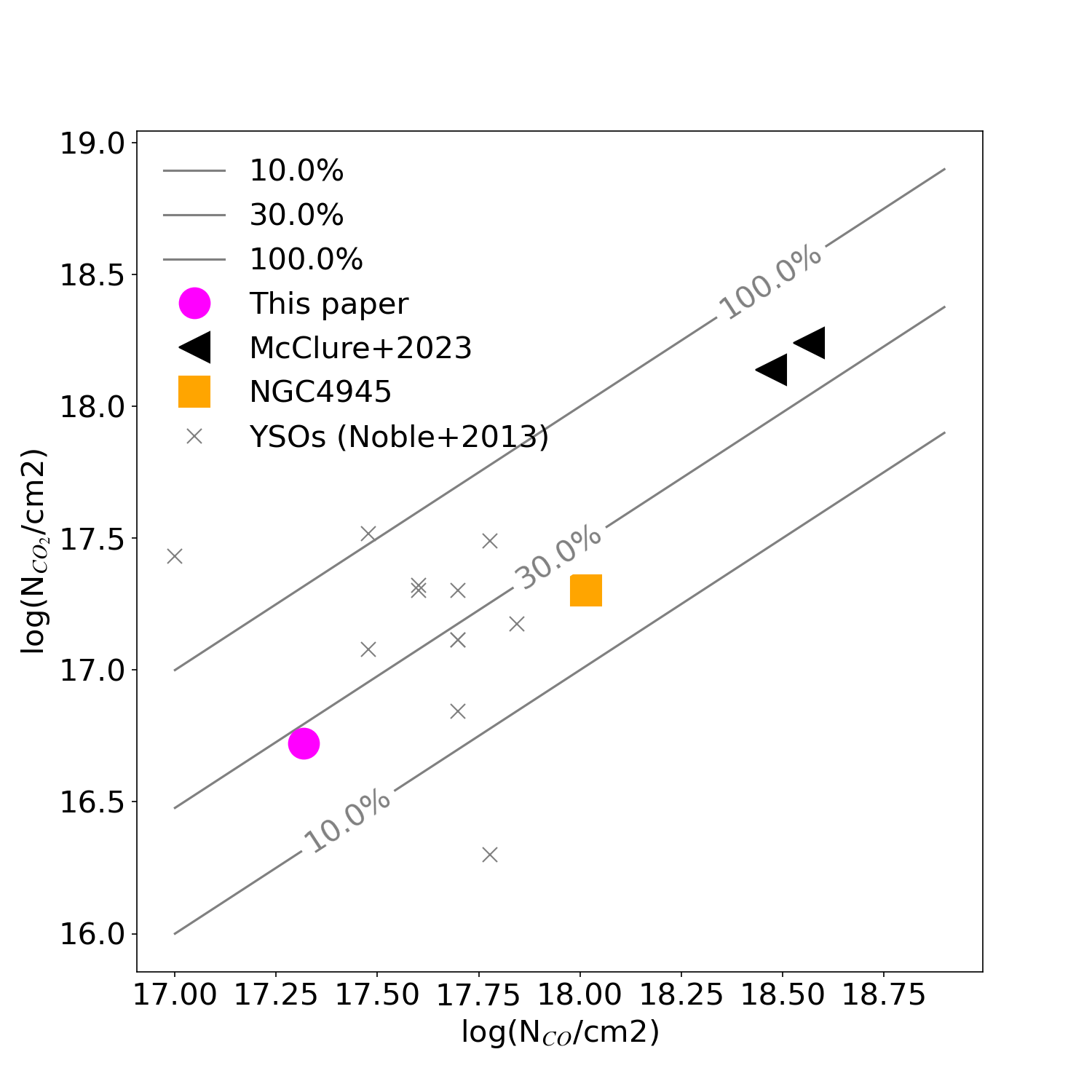}
\caption{A comparison between the column densities in the $^{12}$CO$_2$ and $^{12}$CO ices. We overlay some literature results as indicated. The lines represent the 10$\%$, 30$\%$ and 100$\%$ relations. 
\label{fig:column_density_comparison}}  
\end{figure*}

Figure\,\ref{fig:column_density_comparison} shows how the derived column densities for CO$_2$ vs. CO for FLS2 ($z$=0.494) compare with other sources in the literature including our comparison heavily obscured Galactic objects \citep{McClure2023} as well as YSOs \citep{Noble2013}. We find a remarkably good correspondence such that while our source is on the whole less obscured the relative numbers of CO$_2$ vs. CO molecules in the line of sight are very similar to those comparison Galactic sources with the CO$_2$ molecule abundance in the ice mantles being $\approx$30\% that of the CO molecules on average. We see a very similar CO$_2$ to CO ratio in the buried AGN spectrum of NGC4945. By contrast, while our other extragalactic comparison spectra (NGC7469-ring and VV114-NE) both show clear CO$_2$ absorption, neither have significant solid-phase CO absorption. As discussed in the Introduction ice mantle growth on dust grains is expected to proceed through various stages with $H_20$+CO$_2$ mixture building up first, followed by CO freeze-out at higher densities which can desorb back into the gas-phase as the temperature increases. The CO freeze-out is often accompanied by species including the C$\equiv$N bond as in OCN$^-$, as seen in our source as well as the comparison spectra of NGC4945 and W33 (see Figure\,\ref{fig:opt_depth_co}). Table\,\ref{table:derived_quantities} suggests the CO to OCN$^-$, since we assumed the XCN feature is dominated by OCN$^-$ in deriving the column density, ratio is roughly 5:1. Water ice is the dominant molecule; however, recall that its exact abundance is uncertain due to the complex nature of the 6.0$\mu$m absorption feature as well as the uncertainty in de-blending it from the PAH6.2\,$\mu$m emission feature.  

\begin{deluxetable}{cccccc}
\tabletypesize{\footnotesize}
\tablecaption{Absorption features fit results\tablenotemark{\tiny *} \label{table:derived_quantities}}
\tablehead{
\colhead{Species} & \colhead{$\lambda_{peak}$} & \colhead{FWHM} & \colhead{$\tau_{X}$} & \colhead{$\tau_{X,int}$} & \colhead{$N_X$}\\
 &  \colhead{[$\mu$m]} & \colhead{[$\mu$m]}  & & & \colhead{[/cm$^2$]}}
\startdata
CO$_2$ & 4.2695$\pm$0.0002 & 0.0284$\pm$0.0005 & 0.350$\pm$0.005 & 5.81 & 5.28$\times$10$^{16}$\\
XCN & 4.612$\pm$0.002 & 0.057$\pm$0.003 & 0.19\,$\pm$0.01 & 5.32 & 4.09$\times$10$^{16}$\\
CO & 4.681$\pm$0.002 & 0.036$\pm$0.002 & 0.13$\pm$0.01 & 2.28 & 2.07$\times$10$^{17}$\\
H$_2$0 & 6.076$\pm$0.004 & --  & 0.28 & 41.5 & 3.43$\times$10$^{18}$ \\
\enddata
\tablenotetext{\tiny *}{The quoted uncertainties are statistical based on the Gaussian fits described in Section\,\ref{sec:opt_depth}. These are lower limits on the true uncertainties which also include systematics in the continuum determination as well as ice feature profiles.  }
\end{deluxetable}

\begin{figure*}[h!]
\centering
\includegraphics[scale=0.5,clip=true]{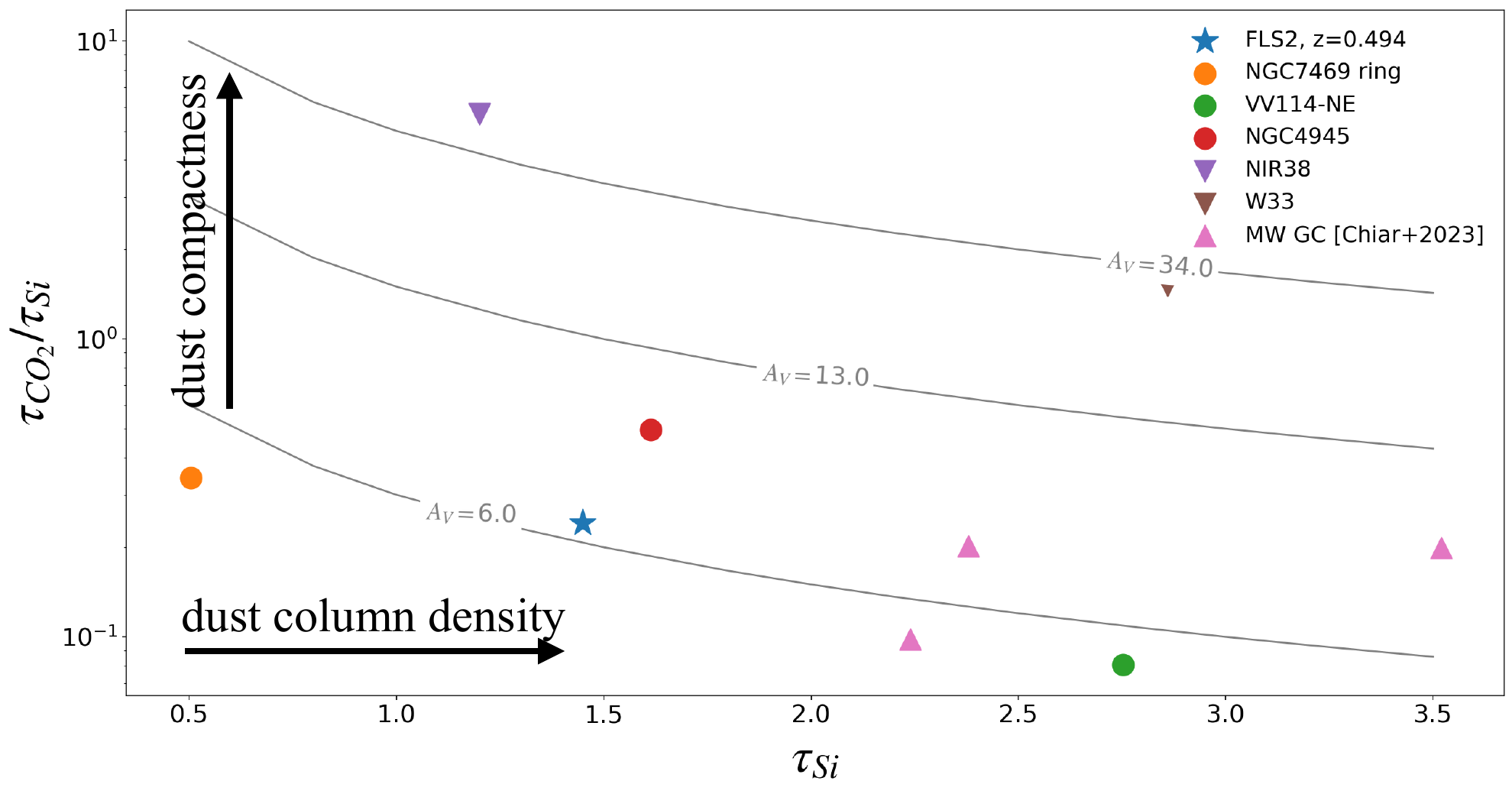}
\caption{A comparison of the optical depths of ice features vs. the silicate 9.7\,$\mu$m feature for FLS2 as well as a variety of comparison Galactic and extragalactic sources as labeled. We overlay lines of constant $A_V$ which are derived from the Galactic $\tau_{CO_2}$ vs. $A_V$ relation \citep{Whittet2007}, hence apply to dense clouds. On the other hand, the $\tau_{Sil}$ may include both diffuse and dense dust. The arrows show the directions of increase of the total dust column density vs. the dust compactness along the line of sight.}
\label{fig:optical_depth_comparison}  
\end{figure*}

\section{Discussion}
\subsection{Evidence for a highly obscured nucleus in SSTXFLS J172458.3+591545 \label{sec:compact_nucleus}}

As discussed in the Introduction, ice mantle build-up requires high physical densities not just high dust column densities. Our sources shows significant CO ice consistent with post CO freeze-out stage which is expected at $n>10^5$\,cm$^{-3}$. This means that we need to have high density gas-dust clumps consistent with the presence of a compact dusty nucleus \citep[see discussion in ][]{Spoon2004,Lai2020}.  The deep silicate feature alone is degenerate between this model and one where the obscuring dust is extended galaxy-wide. Indeed, host galaxy obscuration is observed in the correlations between silicate feature depth and galaxy inclination \citep{Lacy2007,Goulding2012}. This helps explain the lack of correlation between the silicate depth and ice features depth, which is also seen in the significantly variable ratios of the ice to silicate feature depths in Figure\,\ref{fig:compare_goals}.   

Our source's ice absorption features  are quite similar in shape to those seen in the spectrum of the NGC4945 nucleus \citep[see e.g.][]{Perez-Beaupuits2011}, although the ones in our source are about 5\,$\times$ less deep, as seen in Figures\,\ref{fig:opt_depth_co} and Figure\,\ref{fig:column_density_comparison}. NGC4945 is a well studied very nearby, $d\approx3.7$\,Mpc\footnote{Based on several redshift independent distance measures in NED.} buried AGN. The high obscuration ($A_V\approx 14$) in NGC4945 is due in part to it being observed through an edge-on disk. High resolution molecular gas and infrared continuum studies reveal a dense compact nucleus, likely in the form of a circumnuclear disk. There is evidence for both gas outflows and inflows \citep[e.g.][]{Curran2001,Otto2001,Gaspar2022}. 

To determine the levels of dust obscuration in the compact nucleus in FLS2 we use the relationship between $N_{CO_2}$ and $A_V$ from \citet{Whittet2007}. We find an $A_V$ of 6.3-7.3 where the range accounts for the range in $CO_2$ band strength in the literature \citep{Gerakines1995,McClure2023}. These high levels of obscuration account for the fact that while this source shows the unambiguous mid-IR AGN lines [NeV] and [NeVI], it has no clear evidence for an AGN in its rest-frame optical spectrum \citep{Lacy2007}. We note that galaxies with buried nuclei are often considered to host heavily obscured AGN \citep[see][for broader discussion]{Marshall2018}. However, if the obscuration is sufficiently high even the mid-IR AGN lines like [NeV] can become undetectable. In \citet{Spoon2022} (Figure 17), buried nuclei with [NeV] detections  are common until $S_{sil}\approx -1$ to $-1.5$ and largely disappear afterwards. Our source is right in this regime. Any deeper nuclear obscuration is likely to hide even these mid-IR AGN features.  
Using the measured $N_{CO_2}$ to derive $A_V$ for the compact component ignores the possibility of the feature being ``filled-in" by foreground unabsorbed continuum, which is all the more likely considering that the beam within which our spectrum is extracted contains a large portion of the whole galaxy (see Figure\,\ref{fig:full_spectrum}{\it left}), and therefore includes a wide range of environments. Such filling-in from other components of the galaxy within the beam means that what we infer as an optical depth for the nucleus is a lower limit on its true optical depth. For comparison, the spectrum of NGC4945 shown in Figure\,\ref{fig:opt_depth_co} shows 5\,$\times$ deeper features, but represents only the central $\approx 20\times 40$pc of the galaxy \citep{Spoon2004}.  In the above we cannot distinguish between a torus-scale nucleus, i.e. $r\sim10pc$, or in fact a circumnuclear disk as in NGC4945 that we might be seeing edge-on where the size would be more like $r\sim500pc$.

\subsection{The relative strength of ice and silicate features}

As discussed in the Introduction, the depth of the silicate features in galaxies does not correlate with the presence or depth of ice features beyond the fact that ice features only appear in sources that show silicate absorption (at least $\tau_{sil}\approx0.5$). We can consider the silicate feature depth as proxy for the overall column of dust regardless of whether it is distributed throughout the host galaxy or highly concentrated in clumps along the line of sight including a compact, highly obscured nucleus as speculated above. However, clumpy AGN tori models struggle to reproduce the deepest silicate absorption sources \citep[e.g.][]{Nenkova2000}. Sources with $\tau_{sil}>1-2$ require a more smooth dust distribution either as a screen or a shell around a nucleus. They represent the distinct ``buried nucleus" track in the Spoon diagram with PAH equivalent width on one axis and the apparent silicate feature depth on the other \citep[see e.g. ][]{Spoon2007,Marshall2018,Sajina2022}. On the other hand, as argued in Section\,\ref{sec:compact_nucleus}, the stronger the ice features, the higher the physical density which implies more compact structures along the line of sight. 

We explore this model in Figure\,\ref{fig:optical_depth_comparison} which shows the CO$_2$-to-silicate optical depth ratio vs. the silicate feature depth. Here we show both our galaxy along with our two GOALS comparison spectra, NGC4945 and a few literature values from Galactic sightlines. For the nearby buried AGN NGC4945, we extract the optical depth in CO$_2$ from its {\sl Spitzer} IRS spectrum \citep{Perez-Beaupuits2011}. The overlaid lines represents a constant $A_V$ which is translated to constant $\tau_{CO_2}$ via the relation in \citet{Whittet2007}. We then take an array of $\tau_{Si}$ values ranging from 0.5-3.5. The constant $\tau_{CO_2}$ translates to smoothly decreasing $\tau_{CO_2}/\tau_{Si}$ for increasing silicate feature depth. This assumes in essence that the ice mantle build-up is independent of the silicate feature depth (beyond some threshold as observed in galaxies). This plot can be viewed roughly as increasing total dust column density along the x-axis and increasing compactness of the dust along the y-axis. In this context, the labeled $A_V$ values correspond to the $A_V$ associated with the compact dust only, which is the dust associated with ice features. 

This naive model is a reasonably good proxy for most of our comparison extragalactic spectra if we take $A_V\approx6$. In the case of NGC4945, it has an $A_V$ of $\approx$14 towards the circumnuclear disk \citep{Gaspar2022} which is slightly higher than but broadly consistent with this model, especially considering that its $CO_2$ feature detection is from a lower SNR and spatial resolution IRS spectrum. The high $A_V$ Galactic comparison sources (NIR38 and W33A) are also consistent with high $A_V$ here. We remind that the CO$_2$ feature in NIR38 is saturated therefore its depth should be viewed as a lower limit, consistent with its measured $A_V$ being significantly $>30$. However, we stress that we need a lot more especially extragalactic datapoints here to be able to have greater confidence in this model.

Lastly, the three Galactic Center sightlines from \citet{Chiar2002} appear as having $A_V\sim6-10$ here but actually we expect $A_V\approx30$ in the Galactic Center. One explanation might be that much of the column density of dust here arises from us viewing the Galactic Center through the disk of the Milky Way rather than a particularly dense dusty core. The deep silicate absorption features of these sightlines are fully consistent with the canonical $A_V\approx30$ value for the Galactic Center, consistent with our interpretation that the x-axis in Figure\,\ref{fig:optical_depth_comparison} is a rough proxy for the total dust column density. Similarly, using the canonical relation for MW-type dust of $A_V/\tau_{Si}\approx18$, the silicate feature depth of FLS2 implies $A_V\approx25$. This is much larger than the $A_V\approx6$ inferred from the $CO_2$ ice feature alone and is in line with our interpretation that the silicate feature depth is roughly a proxy for the total column of dust whereas the ice features probe only the densest parts thereof -- e.g. a ``dense dusty nucleus". The {\sl HST} imaging of FLS2 also shows an edge-on disk \citep{lacy07_hst} -- see also Eleazer et al. (2025, in prep.).

\subsection{The effect of beam dilution \label{sec:beam_dilution}}
The interpretation of Figure\,\ref{fig:optical_depth_comparison} has an important caveat that we already brought up in Section\,\ref{sec:compact_nucleus}. Our spectrum includes a large fraction of the galaxy -- recall that the MIRI/MRS channel 1 PSF corresponds to a radius of 1.5kpc for our source. This means that both the likely compact dusty nucleus as well as much of the host galaxy are included in the beam. We model this beam dilution using Equation\,\ref{eqn:beam_effect}. 

\begin{equation}
F_{\lambda,model}=bf\times F_{\lambda,nuclear}+(1-bf)\times F_{\lambda,host}
\label{eqn:beam_effect}
\end{equation}

Here $F_{\lambda,model}$ models the observed spectrum around the ice feature (in this case the 4.27\,$\mu$m CO$_2$ feature), whereas $F_{\lambda,nuclear}$ and $F_{\lambda,host}$ represent the separate spectra of the nucleus and host galaxy. The $bf$ parameter represents the fraction of the beam that's filled with the emission from the dusty nucleus vs. the host galaxy. The higher spatial resolution of our data, the larger the $bf$ parameter and vice versa. In the limiting case of only nucleus included in the spectrum, we have $bf=1$ and the host galaxy contribution disappears. Conversely, as $bf$ drops, the relative contribution of the host increases, whereas that of the nucleus decreases.

\begin{figure*}[h!]
\centering
\includegraphics[scale=0.35,clip=true]{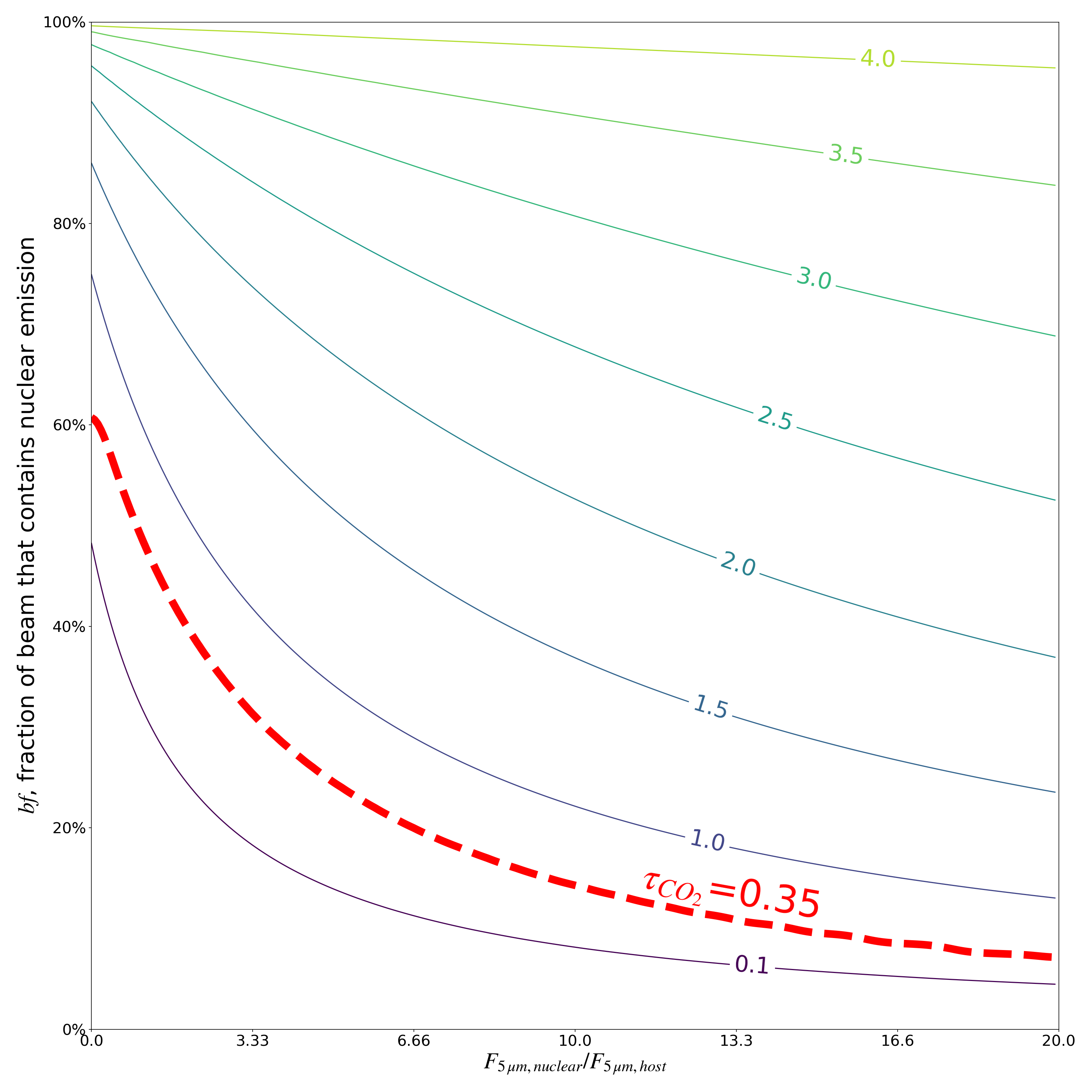}
\caption{An empirical model illustrating the effects of both beam dilution ($y$-axis) and the relative strength of the nuclear vs. host galaxy emission ($x$-axis) on the observed ice feature strengths. The contours are labeled by the observed $\tau_{CO_2}$. See text for details. 
\label{fig:model_tau_co2}}  
\end{figure*}

In Figure\,\ref{fig:model_tau_co2}, we use Equation\,\ref{eqn:beam_effect} to understand the effects on the observed ice feature strength depending both on $bf$ ie what fraction of the beam is due to to the nuclear emission ($y$-axis in Figure\,\ref{fig:model_tau_co2}) and on the relative intrinsic strength of the nuclear vs. host galaxy emission ($x$-axis in Figure\,\ref{fig:model_tau_co2}). To understand the role of this relative strength factor consider that the optical spectra of Type-1 quasars are dominated by the spatially tiny AGN regardless of the spatial resolution of the data because the optical emission of the AGN is much stronger than that of the host galaxy. For the nuclear spectrum we adopt the W33A spectrum (see Figure\,\ref{fig:opt_depth_co2}), one of our most extremely obscured comparison spectra with $\tau_{CO_2}\approx4$. For the host galaxy spectrum, we explored a variety of models but chose to use the spectrum of FLS1 -- a starburst dominated source from our program without any ice features (see Young et al. 2025 in prep.). To minimize noise effects, we fit its 4-5.3\,$\mu$m spectrum with a simple line and use that for $F_{\lambda,host}$. We construct a model grid in terms of our two free parameters: the beam factor $bf$, and the relative strength factor which is $F_{\lambda,nuclear}/F_{\lambda,host}$ evaluated at 5\,$\mu$m. For each point in the grid, we have a model observed spectrum (from Equation\,\ref{eqn:beam_effect}) from which we compute the observed optical depth of the 4.27\,$\mu$m CO$_2$ feature in the same way as we did for FLS2 (see Sections\,\ref{sec:cont_determ}$\&$\ref{sec:opt_depth}). The contours in Figure\,\ref{fig:model_tau_co2} represents equal observed $\tau_{CO_2}$. As expected, at the same relative power of the nuclear emission vs. the host galaxy emission, the observed $\tau_{CO_2}$ decreases with decreasing $bf$ ie as the fraction of the beam that is due to the nucleus decreases. Conversely, at fixed beam dilution factor, $bf$, the smaller the relative power of the nuclear vs. host galaxy emission, the smaller the observed $\tau_{CO_2}$.  

The specific choices of the $F_{\lambda,nuclear}$ and $F_{\lambda,host}$ spectra we adopt, and how we evaluate their relative strength affect the details in Figure\,\ref{fig:model_tau_co2}, but they will not affect the qualitative behavior. 
For example, we explored varying the slope of the $F_{\lambda,host}$ spectrum and found that a steeper spectrum achieves the same observed CO$_2$ feature strength with lower relative nuclear/host fractions than a flatter spectrum. This is unsurprising as the relative strength is evaluated at 5\,$\mu$m which is longward of the CO$_2$ feature. However, the qualitative discussion above remains the same.  

Figure\,\ref{fig:model_tau_co2} highlights the locus of the measured $\tau_{CO_2}$ for FLS2. We can narrow down the parameter space relevant for FLS by estimating the potential fraction of beam covered by the dusty nucleus (i.e. $bf$). Recall that the aperture used here has a radius of 1.5\,kpc (see Section\,\ref{sec:spectra}). If we take the more conservative assumption that the compact nucleus is in the form of a circumnuclear disk whose radius might be $\approx$0.5\,kpc, we estimate a $bf$ of $\approx$10\% ($(0.5/1.5)^2$). This $bf$ estimate is an upper limit as the dusty nucleus could be more compact than assumed here. From Figure\,\ref{fig:model_tau_co2} this low $bf$ implies that the nucleus is likely much more obscured than the measured $A_V\approx 6$, but it is hard to be more precise since much depends on the specific choices in the model. 

Figure\,\ref{fig:model_tau_co2}, with the above estimate for $bf$ also implies an intrinsic $F_{5\mu m,nuclear}/F_{5\,\mu m,host}\approx 15$. This is the intrinsic value before the effects of beam dilution are taken into account. Using the method of \citet{Kirkpatrick2015}, we estimate that only 23\% of the observed mid-IR (5-15\,$\mu$m) light of this source is due to an AGN (see Section,\ref{sec:source}). The Kirkpatrick et al. method relies on the PAH strength divided by the continuum which should scale as $\frac{bf F_{\lambda,nucl}+(1-bf)F_{\lambda,host}}{bf F_{nucl}}$ assuming 100\% of the PAH is due to the host and 100\% of the continuum is from the nuclear or AGN emission. So we can estimate our observed mid-IR AGN fraction as $\frac{(1-bf)}{bf}\frac{F_{host}}{F_{nucl}}$. For FLS2, combining our inferred values for $bf$ and the intrinsic $F_{5\mu m,nuclear}/F_{5\,\mu m,host}$ implies an observed AGN fraction of $\approx 0.9/(0.1*15)\approx 0.6$, evaluated at 5\,$\mu$m. Note that if the $bf$ is actually smaller than estimated, the nuclear to host ratio would be larger leading to approximately the same result. This fraction being larger than the above $\approx$0.2 is because the hotter dust in the vicinity of the AGN means greater dominance of the AGN at 5\,$\mu$m vs. the mid-IR region overall (5-15\,$\mu$m). This rough comparison is reassuring in showing some internal consistency. However, more work is needed to break the degeneracies in AGN strength, obscuration, and spatial resolution by combining diagnostics from both the emission features and absorption features in ``buried nuclei" AGN-star-forming galaxy composites such as FLS2. 

\section{Summary \& Conclusions}
\label{sec:summary}
In this paper we study the ice absorption features in the {\sl JWST} MIRI/MRS spectrum of FLS2, SSTXFLS J172458.3+591545, a $z=0.494$ source observed as part of the {\sl JWST} GO1 program \#1762 ``Halfway to the Peak: A Bridge Program To Map Coeval Star Formation and Supermassive Black Hole Growth" (see Young et al. 2025 in prep for the full survey description).  Here we summarize our findings:
\begin{itemize}
\item We detect the 4.27\,$\mu$m $^{12}$CO$_2$, the 4.62\,$\mu$m XCN/$OCN^-$, the 4.67\,$\mu$m $^{12}$CO and the 6.0\,$\mu$m water ice absorption features. All ice feature profiles have analogs among known Galactic or extragalactic obscured sources.  
\item The relative strength of the CO$_2$ and CO ice features is similar to that of much more highly obscured Galactic sources with the CO$_2$ column density being roughly 30\% that of CO ice -- suggesting similar patterns of ice mantle growth across a wide range of obscuration.  
\item The presence of significant solid phase CO absorption with only hints of gas-phase CO is consistent with high density ($n>10^5 cm^{-3}$) and low temperatures of $T\approx20-90K$. 
\item The above high densities imply the presence of a compact dusty nucleus as opposed to a model where all the dust is distributed smoothly across the host galaxy. We estimate an $A_V\approx6-7$ for the optical extinction towards this nucleus. Beam dilution effects mean this is a lower limit on the true nuclear optical depth. 
\item A plot of the observed $\tau_{Si}$ vs. $\tau_{CO_2}/\tau_{Si}$ is a rough proxy for the total dust column density on the x-axis vs. the level of compactness of this dust on the y-axis. However, the limited beam covering fraction of the suggested compact dusty nucleus means the measured ice features depth is a lower limit. 

\end{itemize}

All the {\it JWST} data used in this paper can be found in MAST: \dataset[10.17909/xa5q-6z41]{http://dx.doi.org/10.17909/xa5q-6z41}.

\section*{Acknowledgments}

We are grateful to the anonymous referee for their careful reading of the paper and detailed feedback which improved significantly the clarity of the paper. We are also grateful to Emmanuel Dartois for useful discussions. This work is based on observations made with the NASA/ESA/CSA James Webb Space Telescope. The data were obtained from the Mikulski Archive for Space Telescopes at the Space Telescope Science Institute, which is operated by the Association of Universities for Research in Astronomy, Inc., under NASA contract NAS 5-03127 for JWST. These observations are associated with {\sl JWST} GO PID\#1762. This work is supported by STSci grant JWST-GO-01762.010-A.  This research has made use of the NASA/IPAC Extragalactic Database (NED),
which is operated by the Jet Propulsion Laboratory, California Institute of Technology,
under contract with the National Aeronautics and Space Administration.

\software{Astropy \citep{astropy:2013, astropy:2018, astropy:2022}} 

\bibliography{midir_spectra}

\end{document}